\documentclass[usegraphicx]{mn2e}

\usepackage{times}
\usepackage{amssymb}
\usepackage{amsmath}
\begin{document}


\newcommand{\Mpc}{\rm\thinspace Mpc}
\newcommand{\kpc}{\rm\thinspace kpc}
\newcommand{\pc}{\rm\thinspace pc}
\newcommand{\km}{\rm\thinspace km}
\newcommand{\m}{\rm\thinspace m}
\newcommand{\cm}{\rm\thinspace cm}
\newcommand{\cmps}{\hbox{$\cm\s^{-1}\,$}}
\newcommand{\cmpssq}{\hbox{$\cm\s^{-2}\,$}}
\newcommand{\cmsq}{\hbox{$\cm^2\,$}}
\newcommand{\cmcu}{\hbox{$\cm^3\,$}}
\newcommand{\pcmcu}{\hbox{$\cm^{-3}\,$}}
\newcommand{\pcmcuK}{\hbox{$\cm^{-3}\K\,$}}

\newcommand{\yr}{\rm\thinspace yr}
\newcommand{\Gyr}{\rm\thinspace Gyr}
\newcommand{\s}{\rm\thinspace s}
\newcommand{\ks}{\rm\thinspace ks}

\newcommand{\GHz}{\rm\thinspace GHz}
\newcommand{\MHz}{\rm\thinspace MHz}
\newcommand{\Hz}{\rm\thinspace Hz}

\newcommand{\K}{\rm\thinspace K}

\newcommand{\Kpcmc}{\hbox{$\K\cm^{-3}\,$}}

\newcommand{\g}{\rm\thinspace g}
\newcommand{\gpcm}{\hbox{$\g\cm^{-3}\,$}}
\newcommand{\gpcmps}{\hbox{$\g\cm^{-3}\s^{-1}\,$}}
\newcommand{\gps}{\hbox{$\g\s^{-1}\,$}}
\newcommand{\Msun}{\hbox{$\rm\thinspace M_{\odot}$}}
\newcommand{\Msunpc}{\hbox{$\Msun\pc^{-3}\,$}}
\newcommand{\Msunpkpc}{\hbox{$\Msun\kpc^{-1}\,$}}
\newcommand{\Msunppc}{\hbox{$\Msun\pc^{-3}\,$}}
\newcommand{\Msunppcpyr}{\hbox{$\Msun\pc^{-3}\yr^{-1}\,$}}
\newcommand{\Msunpyr}{\hbox{$\Msun\yr^{-1}\,$}}

\newcommand{\MeV}{\rm\thinspace MeV}
\newcommand{\keV}{\rm\thinspace keV}
\newcommand{\eV}{\rm\thinspace eV}
\newcommand{\erg}{\rm\thinspace erg}
\newcommand{\Jy}{\rm Jy}
\newcommand{\ergpcmc}{\hbox{$\erg\cm^{-3}\,$}}
\newcommand{\ergcmcups}{\hbox{$\erg\cm^3\ps\,$}}
\newcommand{\ergpcmps}{\hbox{$\erg\cm^{-3}\s^{-1}\,$}}
\newcommand{\ergpcmsqps}{\hbox{$\erg\cm^{-2}\s^{-1}\,$}}
\newcommand{\ergpcmsqpspA}{\hbox{$\erg\cm^{-2}\s^{-1}$\AA$^{-1}\,$}}
\newcommand{\ergpcmsqpspsr}{\hbox{$\erg\cm^{-2}\s^{-1}\sr^{-1}\,$}}
\newcommand{\ergpcmcups}{\hbox{$\erg\cm^{-3}\s^{-1}\,$}}
\newcommand{\ergps}{\hbox{$\erg\s^{-1}\,$}}
\newcommand{\ergpspmp}{\hbox{$\erg\s^{-1}\Mpc^{-3}\,$}}
\newcommand{\keVpcmsqpspsr}{\hbox{$\keV\cm^{-2}\s^{-1}\sr^{-1}\,$}}

\newcommand{\dyn}{\rm\thinspace dyn}
\newcommand{\dynpcmsq}{\hbox{$\dyn\cm^{-2}\,$}}

\newcommand{\kmps}{\hbox{$\km\s^{-1}\,$}}
\newcommand{\kmpspmp}{\hbox{$\km\s^{-1}\Mpc{-1}\,$}}
\newcommand{\kmpspMpc}{\hbox{$\kmps\Mpc^{-1}$}}

\newcommand{\Lsun}{\hbox{$\rm\thinspace L_{\odot}$}}
\newcommand{\Lsunppc}{\hbox{$\Lsun\pc^{-3}\,$}}

\newcommand{\Zsun}{\hbox{$\rm\thinspace Z_{\odot}$}}
\newcommand{\gauss}{\rm\thinspace gauss}
\newcommand{\arcsecond}{\rm\thinspace arcsec}
\newcommand{\chisq}{\hbox{$\chi^2$}}
\newcommand{\delchi}{\hbox{$\Delta\chi$}}
\newcommand{\ph}{\rm\thinspace ph}
\newcommand{\sr}{\rm\thinspace sr}

\newcommand{\pccm}{\hbox{$\cm^{-3}\,$}}
\newcommand{\psqcm}{\hbox{$\cm^{-2}\,$}}
\newcommand{\pcmsq}{\hbox{$\cm^{-2}\,$}}
\newcommand{\pmpc}{\hbox{$\Mpc^{-1}\,$}}
\newcommand{\pmpccu}{\hbox{$\Mpc^{-3}\,$}}
\newcommand{\ps}{\hbox{$\s^{-1}\,$}}
\newcommand{\pHz}{\hbox{$\Hz^{-1}\,$}}
\newcommand{\pcmK}{\hbox{$\cm^{-3}\K$}}
\newcommand{\phpcmsqps}{\hbox{$\ph\cm^{-2}\s^{-1}\,$}}
\newcommand{\psr}{\hbox{$\sr^{-1}\,$}}
\newcommand{\pspsqas}{\hbox{$\s^{-1}\,\arcsecond^{-2}\,$}}

\newcommand{\ergpspcmpK}{\hbox{$\erg\s^{-1}\cm^{-1}\K^{-1}\,$}}

\title{Conduction and cooling flows}
\author[]
{\parbox[]{6.in} {L. M. Voigt, R. W. Schmidt, A. C. Fabian,
    S. W. Allen and R. M. Johnstone\\
\footnotesize
Institute of Astronomy, Madingley Road, Cambridge CB3 0HA\\
}}

\maketitle
\begin{abstract}
Chandra and XMM-Newton observations have confirmed the presence of large
temperature gradients within the cores of many relaxed clusters of galaxies. Here
we investigate whether thermal conduction operating over those gradients
can supply sufficient heat to offset radiative cooling.  Narayan \&
Medvedev (2001) and Gruzinov (2002) have
noted, using published results on cluster temperatures, that conduction within a factor of a few of the Spitzer rate is
sufficient to balance bremsstrahlung cooling. From a
detailed study of the temperature and emission measure profiles of Abell 2199 and
Abell 1835, we find that the heat flux required by conduction is
consistent with or below the rate predicted by Spitzer in the outer regions of the core. Conduction may therefore explain the lack of
observational evidence for large mass cooling rates inferred from
arguments based
simply on radiative cooling, provided that conductivity is suppressed by
no more than a factor of three below the full Spitzer rate. To stem cooling in the cluster centre, however, would necessitate conductivity
values at least a factor of two larger than the Spitzer values, which we
consider implausible. This may provide an explanation for the
observed star formation and optical nebulosities in cluster cores.
The solution is likely to be time dependent. We briefly discuss the
possible origin of the cooler gas and the implications for massive
galaxies.
\end{abstract}


\section{Introduction}

The intracluster gas at the centres of many clusters of galaxies has a
radiative cooling time of a billion years or less (Peres et al. 1997),
suggesting that cooling flows may be operating there (Fabian 1994 and
references therein). Chandra has now enabled spatially detailed
temperature profiles to be made of such cluster cores and large
temperature drops, by up to a factor of three, are commonly seen
(Allen et al. 2001a,b,c; Schmidt et al. 2001; Ettori et al. 2002;
Johnstone et al. 2002; David et al. 2001; McNamara et al. 2001;
Molendi et al. 2001). Results from the high spectral resolution
Reflection Grating Spectrometer (RGS) on XMM-Newton also show temperature components down to about one third of
the overall cluster virial temperature (Peterson et al. 2001; Tamura et
al. 2001; Kaastra et al. 2001; Peterson 2002). It is of great importance,
however, that RGS spectra do not show evidence for lines
expected from gas cooling at lower temperatures (e.g.
from Fe XVII).

This general result means that a simple cooling flow is not operating
in cluster cores and has provoked the study of a wide range of
possibilities, including absorption of the soft X-rays to mixing or
heating of the cooler gas (Peterson et al. 2001; Fabian et al. 2001,
2002; Churazov et al. 2001; B\"ohringer et al. 2002; Br\"uggen \& Kaiser
2001). If heating is the answer then it has to be distributed over
radius and temperature and is not just concentrated at the centre
(Johnstone et al. 2002).

Another process which has long been discussed for clusters is thermal
conduction ( see e.g. Takahara \& Takahara 1979; Tucker \& Rosner
1983; Friaca 1986; B\"ohringer \& Fabian 1989; Bertschinger \& Meiksin
1986; Bregman \& David 1988; Sparks 1992; Narayan \& Medvedev
2001; Suginohara \& Ostriker 1998). For a multiphase cooling flow to
operate conduction has to be suppressed (Binney \& Cowie 1981; Fabian 1994; see Chandran \& Cowley 1998; Pistinner, Levinson \&
Eichler 1996; Malyshkin 2001 for possible mechanisms). Some
observational evidence for highly suppressed conduction (Ettori \&
Fabian 2000) comes from the `cold fronts' found in Chandra images of many
clusters (Markevitch et al. 2000; Vikhlinin et al. 2001; Mazzotta et al.
2002). The serious problem relevant to cooling flows here is that
conduction works best at high temperatures and cooling at low
temperatures, so it is difficult to see how they can balance throughout the
(observed) range of temperature (Bregman \& David 1988; Fabian,
Canizares \& B\"ohringer 1994). 

More explicitly, for a spherical, stationary, constant pressure, cooling flow the energy equation is 

\begin{equation}
-{\dot M\over{4 \pi r^2}}
{d\over{dr}}{{5kT}\over{2\mu}}=-n^2\Lambda +{{1\over
r^2}{\partial\over{\partial r}}\left(r^2
\kappa {\partial T\over {\partial r}}\right)} 
\end{equation}
where $n,T,\mu,\Lambda$ and $\kappa$ are the gas density, temperature,
mean molecular weight, cooling function and thermal conductivity,
respectively. If there is no flow ($\dot M$ = 0) and $\kappa \propto
\kappa_{s}$, the Spitzer conductivity ($\kappa_{s}\propto T^{5/2}$),
then since $\Lambda \propto T^\alpha$, $T\propto r^{2/(11/2-\alpha)}$
(Fabian et al. 1994). So, where bremsstrahlung cooling dominates
(above $T\sim 2\keV$) and $\alpha\sim0.5$, then $T\propto r^{0.4}$
(below $T\sim 1\keV$, where line cooling dominates and $\alpha\sim -0.5$,
then $T\propto r^{0.3}$). Such a gradient is similar to those found
from the Chandra images (e.g.  Johnstone et al. 2002). This in no way
proves that conduction operates but encourages further investigation.
Narayan \& Medvedev (2001) and Gruzinov (2002) have recently noted,
using published results on cluster temperatures, that conduction
suppressed by a factor of a few below the Spitzer rate will have a
significant effect on cluster cooling flows.

Here we carry out a detailed study of the heat influx needed to
balance the radiative energy loss in a low temperature cluster (Abell 2199;
Johnstone et al. 2002) and a high temperature one (Abell 1835; Schmidt et al.
2001). Rather than test the hypothesis that thermal conduction is
sufficient by looking at the temperature gradient at just one
radius, we compute the losses and fluxes as a function of temperature
and radius, as first carried out by Stewart et al. (1984). Our results
show that the heat influx required to balance radiation losses occurs
at a rate which is consistent with or below the
Spitzer (1962) relation in the outer parts of the core (where the core
is taken as the region inside the cooling radius of the cluster), but at a rate
above that given by Spitzer in the very centres of the clusters. We
also find that if conduction is suppressed by more than a factor of
three below
the Spitzer rate then heat transfer from
hotter gas at large radii is unable to stop a cooling flow from developing.

If conduction is the solution to the problem of cooling flows, then it
begs the question of how the cooler gas components originate. If the
presence of cooler gas is due to radiative cooling of the hotter gas, then it is not obvious
how such cooling has been arrested. Possibly it is due to the
introduction of cooler subclusters, the dense cores of which are not
thoroughly shocked (Fabian \& Daines 1991; Motl et al. 2001). 

\section{The temperature and emission measure profiles used}

Abell 2199 and Abell 1835 are relaxed clusters at luminosity distances
of 187 Mpc ($z$=0.0309) and 1599 Mpc ($z$=0.2523), respectively. We assume a
cosmology with $H_{0}$=50 $\kmpspMpc$ and $q_{0}$=0.5 throughout.

\begin{figure*}
\centering
\includegraphics[angle=-90, width=0.95\columnwidth]{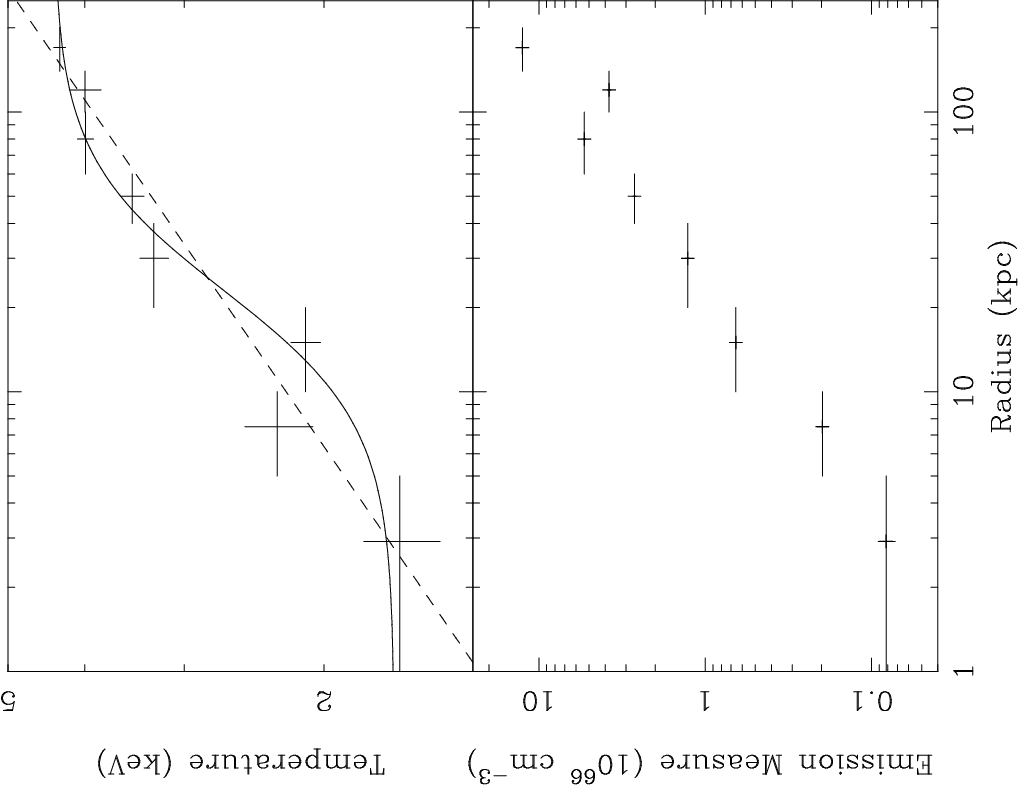}
\hspace{5.0mm}
\includegraphics[angle=-90, width=0.95\columnwidth]{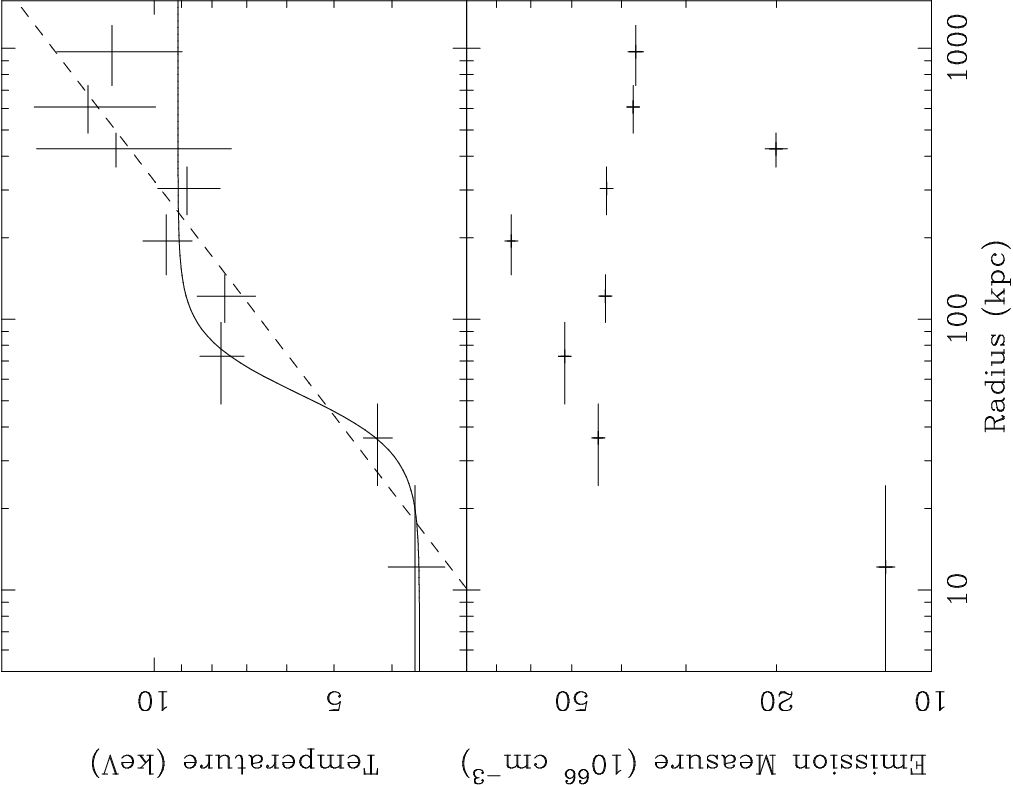}

\caption{The observed (deprojected) spectrally-determined temperature
  and emission measure profiles for Abell 2199 (left; Johnstone et
  al. 2002) and Abell 1835 (right; Schmidt et al. 2001). The one sided error
 bars plotted (necessary for chi-square calculations) are the root-mean-square of the two sided error bars
  given in the aforementioned papers. The dashed
  lines are the best fit power laws to Abell 2199 ($\chi^{2}$=39.5, 6
 degrees of freedom) and Abell 1835 ($\chi^{2}$=16.4, 7
 degrees of freedom). The solid lines are the best fit curves to Abell
  2199
 ($\chi^{2}$=12.3, 4
 degrees of freedom) and Abell 1835 ($\chi^{2}$=5.5, 5
 degrees of freedom) using the functional form in Equation (6).}
\label{fig:emtfits}
\end{figure*}

Deprojected spectrally-determined temperature and density profiles
obtained for Abell 2199 by Johnstone et al. (2002) and for Abell 1835
by Schmidt et al. (2001) were used to carry out the calculation.
Briefly, photon samples extracted from annuli (eight for Abell 2199
and nine for Abell 1835) around the cluster centre were used to obtain
a single (deprojected) temperature and density for each spherical shell by
fitting the spectral energy distribution of consecutive annuli with multiple MEKAL (Kaastra \& Mewe 1993) plasma models (one mekal model for the
outermost annulus, two for the next one in, etc.) absorbed by the PHABS
photoelectric absorption model (Balucinska-Church \& McCammon
1992). The fitting was carried out using the PROJCT routine in XSPEC
(Arnaud 1996). A more detailed
description of the deprojection technique is given by Allen et
al. (2001b). We note that this approach gives similar results to that of Ettori et
al. (2002) when tested on the same cluster. The
temperature and emission measure ($EM$=$n^{2}$ $\times$ $shell$ $volume$) profiles for each cluster are shown in
Figure~\ref{fig:emtfits}. 

\section{The derivation of $\kappa$}

The conductivities required to balance radiative losses by conduction of
heat from the outer, hotter parts of the clusters into the central
regions, where the cooling time is less than the Hubble time, are
calculated using the temperature and emission measure profiles shown
in Figure~\ref{fig:emtfits}. Setting $\dot M$ = 0 in Equation (1)
and integrating gives

\begin{equation}
\int\limits_{\mathcal{V}} n^{2} \Lambda(T)\hspace{0.7mm}\mathrm{d}\mathcal{V} =
\int\limits_{\mathcal{S}} \kappa (\boldsymbol{\nabla} T).\mathrm{d}\mathcal{\boldsymbol{S}}
\end{equation}
where $\mathcal{V}$ is the volume and $\mathcal{S}$ the surface area of the X-ray emitting sphere.

As the data are binned into $m$ shells, the above equation is approximated
to the following sum in order to calculate the conductivity
coefficient, $\kappa_{j}$, at the outer boundary of the $j$th shell,
\\
\\
For  $j$: \{1 $\leq$ $j$ $\leq$ $m$\} 
\begin{equation}
\sum_{i=1}^{j}EM_{i}\Lambda_{i}=\kappa_{j}\times\left(\frac{\Delta
    T}{\Delta r}\right)_{(j+1)-j}\times 4 \pi r_{j}^{2}
\end{equation}
where $EM_{i}\Lambda_{i}$ is the rate of energy loss from the $i$th shell, $r_{j}$
is the radial distance to the outer boundary of the $j$th shell and
$\Delta T_{(j+1)-j}$ and $\Delta r_{(j+1)-j}$ are the
temperature difference and radial distance between the centres of
consecutive shells, respectively.

Models were fitted to the temperature profiles in order to calculate the temperature
gradient between shells. This ensures that the temperature always decreases towards
the cluster centre and enables us to handle the
uncertainties in our calculation. Taking the temperature difference between
consecutive data points does not always provide meaningful results since
several of the error bars overlap, allowing the possibility of negative temperature gradients and zero
gradients, the latter consequence leading to infinite error bars on
the computed conductivity values.   

The cooling function was evaluated using MEKAL (with an abundance
relative to solar of 0.4) and the conductivity values computed using
Equation (3) plotted against
the temperature given by the model at $r_{j}$. Only shells lying within the cooling radius of the cluster were
used to compute $\kappa$; $r_{cool}$
$\sim$ 150kpc for Abell 2199 (Johnstone et al. 2002) and $r_{cool}$ $\sim$
200kpc for Abell 1835 (Schmidt et al. 2001). The results were compared with the conductivity--temperature
relation given by Spitzer (1962). 

The thermal conductivity of a completely ionized gas in a space free
from magnetic fields and for which the ions are all of one kind
(hydrogen) is given by Spitzer (1962) as

\begin{equation}
\kappa_{s} = \frac{1.84\times10^{-5} T^{\frac{5}{2}}}{\mathrm{ln}\hspace{0.6mm}\Lambda} \ergpspcmpK
\end{equation}
where ln\hspace{0.6mm}$\Lambda$ is the Coulomb logarithm (see Cowie \& McKee
1977). The accuracy of this equation does not exceed 5-10 per cent
(see Spitzer \& H\"arm 1953). For electron densities and temperatures appropriate to the
clusters considered, ln\hspace{0.6mm}$\Lambda$ $\sim$ 37, and

\begin{equation}
\kappa_{s} \simeq 5.0\times10^{-7} T^{\frac{5}{2}} \ergpspcmpK
\end{equation}
In a plasma with magnetic fields the conductivity may be reduced by a
factor of three or more below the Spitzer value (Narayan \& Medvedev
2001; Malyshkin 2001).

\begin{figure*}
\centering
\includegraphics[angle=-90, width=0.95\columnwidth]{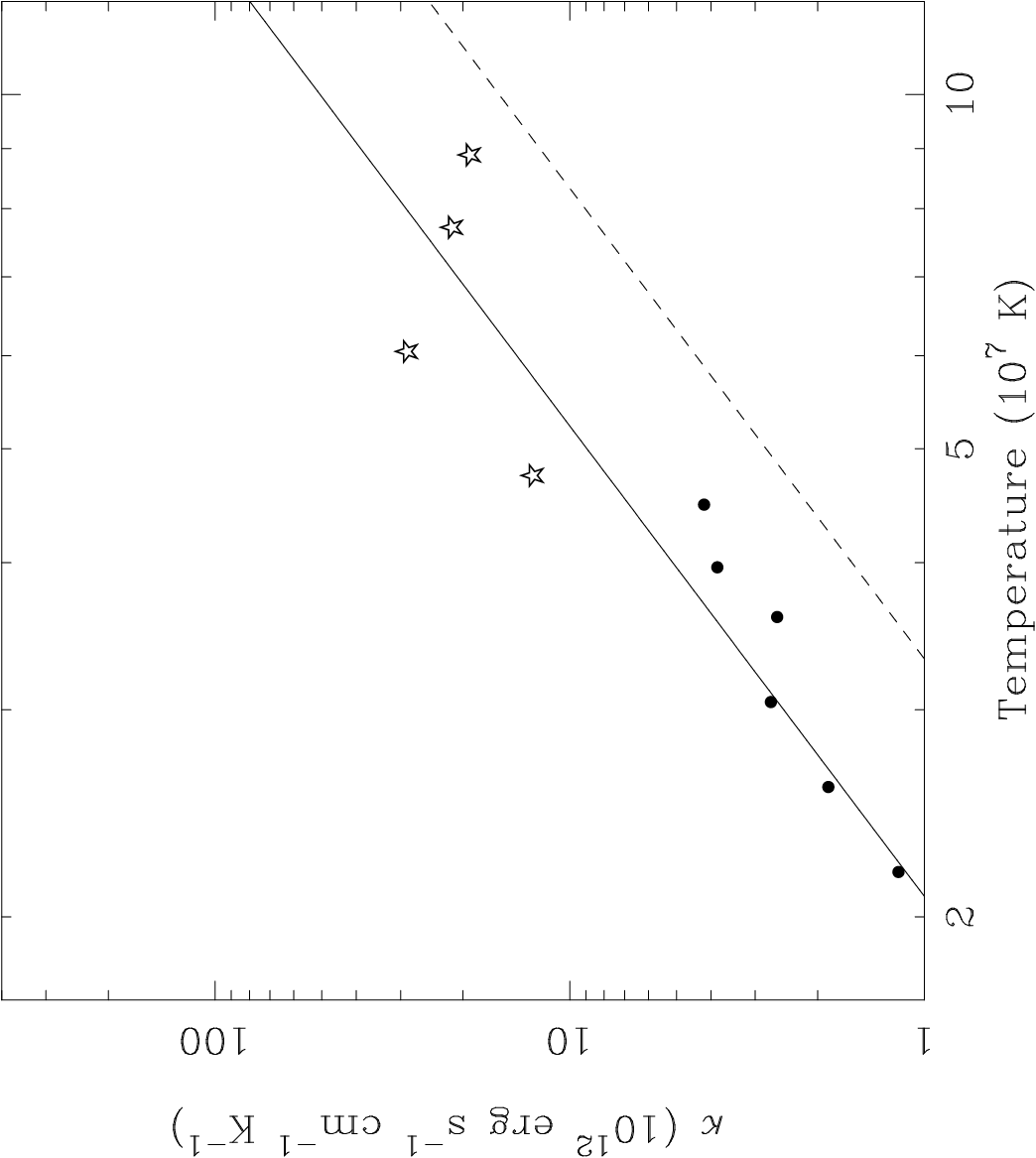}
\hspace{5.0mm}
\includegraphics[angle=-90, width=0.95\columnwidth]{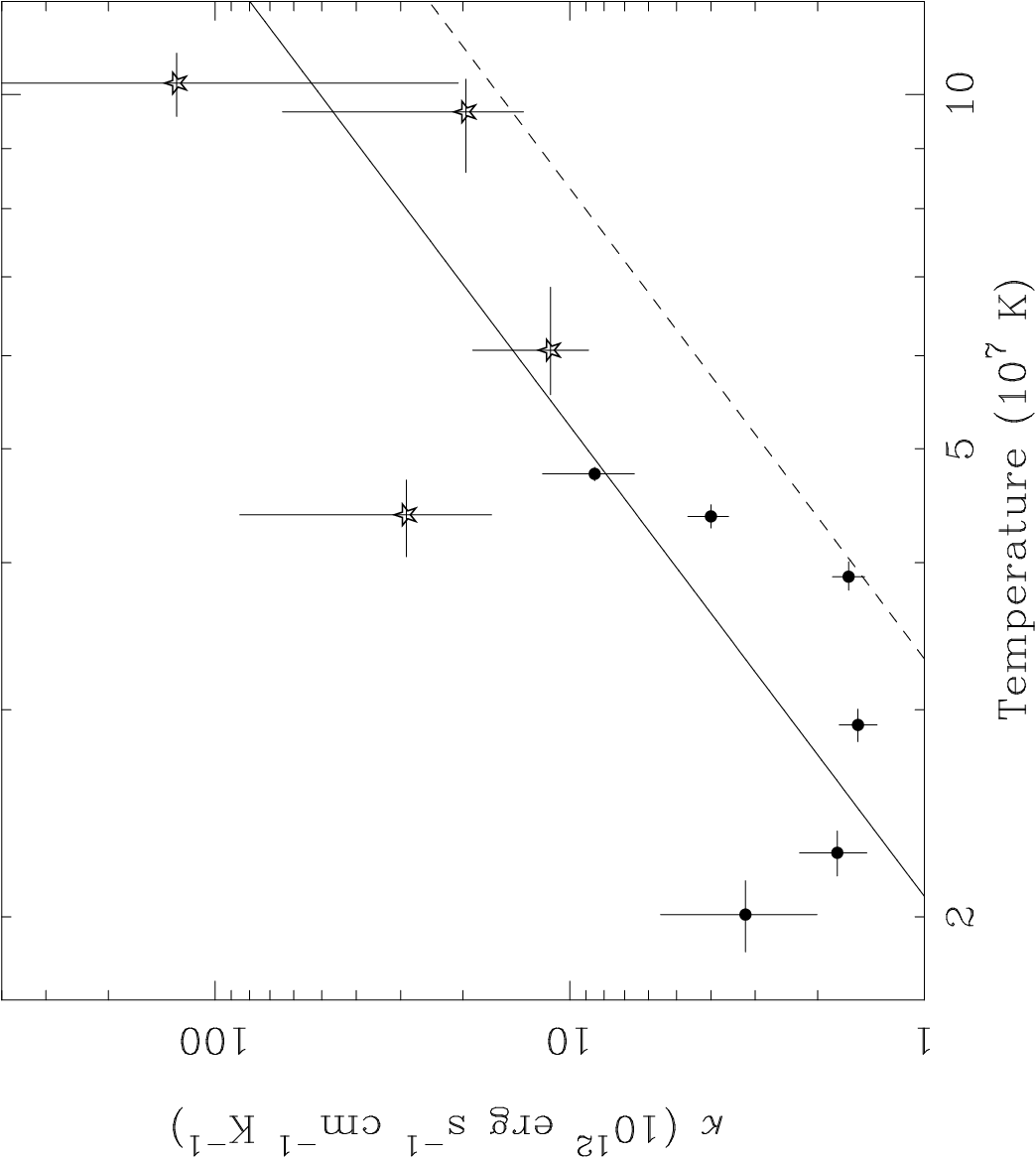}
\caption{The conductivity coefficients required for
  conduction to balance radiation loss using power law fits (left) and
  using the functional form given in Equation (6) (right) to the
  temperature profiles. Filled circles represent the results for Abell 2199
  and open stars for Abell 1835. The upper error
  bar on the fourth data point for Abell 1835 is $4.9 \times 10^{15}$
  $\ergpspcmpK$. The solid line is the Spitzer conductivity and
  the dashed line is one third of the Spitzer conductivity.}
\label{fig:spitzer}
\end{figure*}

Simple power law fits to the deprojected data are shown in Figure
 \ref{fig:emtfits} ($\chi^{2}_{\nu}$ $\sim$ 6.6 for Abell 2199 and $\chi^{2}_{\nu}$
$\sim$ 2.3 for Abell 1835). Using
these models we find conductivity values
which scatter around the Spitzer curve (see Figure
\ref{fig:spitzer}). However, the reduced chi-square value of the fit
 to the temperature profile of Abell 2199 is unacceptably high. We
therefore looked for a model which provided a better reduced
chi-square for both cluster profiles.

The temperature profiles were modelled using a function of the form

\begin{equation}
T_{r} = T_{0} +
T_{1}\left[\frac{\left(\frac{r}{r_{c}}\right)^{\eta}}{1+\left(\frac{r}{rc}\right)^{\eta}}\right]
\end{equation}
where $T_{r}$ is the temperature at a radial distance $r$ from the cluster
centre. Allen et al. (2001b) use this model as a reasonable universal fit to
projected temperature profiles of six cooling flow
clusters, including Abell 1835. We adopt it here for the deprojected
temperature profiles as a smooth fitting function. $T_{0}$, $T_{1}$, $r_{c}$ and $\eta$ are free parameters determined using chi-square fitting. 

The best fit models to the temperature data are shown in Figure~\ref{fig:emtfits}.  For Abell 2199, $\chi_{\nu}^{2}$=3.1
($kT_{0}$=1.6 keV, $kT_{1}$=2.7 keV, $r_{c}$=30.1 kpc, $\eta$=1.8) and
for Abell 1835, $\chi_{\nu}^{2}$=1.1 ($kT_{0}$=3.6 keV,
$kT_{1}$=5.5 keV, $r_{c}$=59.3 kpc, $\eta$=4.1). The probabilities of
exceeding the chi-square values are 0.02 and 0.35 for Abell 2199 and
Abell 1835, respectively. Although the fit to Abell 2199 is only
marginally acceptable, this
 function provides much improved reduced chi-square values over the
power law fits.
 
The physical motivation for using these models must also be considered. Both clusters
contain radio sources and as such are not expected to have temperature
distributions which are smoothly increasing functions of
radius. It is likely that the profiles are
time-dependent, undergoing fluctuations from a smooth curve during
radio outbursts lasting $\sim$ 10$^{7}$ years. Since
conduction must operate on timescales $\sim$ 10$^{9}$ years in order to
successfully offset cooling, this calculation requires the
time-averaged cluster profiles. We reason that as the functional form
used here is a fair representation of several cooling flow
clusters, it is a plausible time-averaged model. It is certainly more
physically realistic than the power law models since it has a vanishing gradient
both at large and small radii. We proceed using this function and take
caution when analysing the results for Abell 2199.   

Conductivity values and corresponding 1$\sigma$ uncertainties were found using Monte Carlo
simulations. Ten thousand random temperature profiles consistent with the
data were generated and fitted using a chi-square
minimization routine. The median conductivity and
temperature value at each radius are plotted in Figure~\ref{fig:spitzer},
with the central 68 per cent
of each distribution providing the error bars. The emission measure
errors are negligible in comparison.

In both clusters we find conductivity values which are less than a factor
of three below
the Spitzer curve in the outer regions and which lie above the curve
in the innermost regions. The latter result shows that conduction at
the full Spitzer rate is insufficient to balance
cooling in the centres of Abell 2199 and Abell 1835. Whether the higher conductivity required for the hottest,
outermost points is important depends on the age of the cluster.
We used the temperature difference given by the model between shell centres, rather than the derivative at the
outer boundary of each shell, so as not to rely on the model too
rigorously. It is reassuring that simple power law fits
to the data confirm the requirement for conduction at a rate greater
than the Spitzer rate in the lowest temperature regions.

\section{Discussion}

We have shown for Abell 2199 and Abell 1835 that heat conduction at a
rate close to the Spitzer rate is sufficient to balance cooling in
the outer regions of the core. However, if conductivity is suppressed
below the Spitzer rate by more than a factor of three, as is possible in a plasma with a
tangled magnetic field, heating by conduction merely reduces, but
cannot prevent cooling. This result is clearly dependent on the value of
$h$ = $H_{0}/100 \kmpspMpc$, since
$\kappa$\hspace{0.6mm}$\propto$\hspace{0.6mm}$h^{-1}$. Thus the lowest
value we find for $\kappa$ at about one third the Spitzer rate drops to one
fifth that rate if $h = 0.75$. (In principle, if we can accurately
predict the value of $\kappa$ and can assume that conduction balances
radiative cooling then we have a new method for determining $H_0$).

Towards the innermost regions conduction at the
full Spitzer rate is unable to offset cooling, implying that, in the absence
of an additional heat source, a small cooling flow will operate in the
central (10-20 kpc), cooler
parts of the cluster cores. We consider this likely, given the possibility
that the solutions to the energy equation (1) are unstable (Fabian et al.
1994). Observationally, many of the Chandra and RGS limits allow for a
small but significant flow (e.g. McNamara et al. 2000; David et al.
2001; Allen et al. 2001a,b; Peterson et al. 2001; Peterson 2002),
which can supply the observed star formation and optical nebulosities
(Johnstone, Fabian \& Nulsen 1987; Heckman et al. 1989; Allen 1995; Cardiel et al.
1998; Crawford et al. 1999) and cold gas (Edge 2001 and references
therein). We note that the radius of the inner region where the
required conductivity exceeds the Spitzer value coincides with the
apparent break in the $\dot M$ profiles found from a cooling flow
analysis of the data (Schmidt et al. 2001;  Johnstone et al. 2002).

Given the opposing temperature dependences of conduction and cooling,
further work is required to show whether initially isothermal regions
can develop the
large, central temperature gradients observed. The cooler gas may be a
remnant of earlier subclusters, the dense cool gas cores of which have
only partially merged with the main cluster. The situation is
time-dependent. The cold fronts first
reported by Markevitch et al. (2000) and the transonic motion of some
of the gas (Vikhlinin et al. 2001) are evidence for this scenario. 
Presumably conduction is highly suppressed across a moving front, due to
organised magnetic fields, but may rapidly set in at the rear when the gas is decelerated. A radial field will, of course, be
optimal for conduction. Such a field could be the consequence of a
cooling inflow (Bregman \& David 1988).

Much further work on a wider range of clusters is required to test the
conduction hypothesis. It will be most interesting if a simple
physical process such as conduction is responsible for limiting the
total cooled gaseous mass of the largest galaxies. It will also be
interesting to investigate whether conduction is operating in the
interstellar medium of elliptical galaxies (Saito \& Shigeyama 1999), thereby explaining why
they have very short radiative cooling times at their centres, yet
little evidence for cooling flows.

\section{Acknowledgements}
LMV acknowledges support from PPARC.
ACF \& SWA thank the Royal Society for support.


\begin{thebibliography}{}

\bibitem []{} Allen S.~W., 1995, MNRAS, 276, 947

\bibitem []{} Allen S.~W. et al., 2001a, MNRAS, 324, 842
\bibitem []{} Allen S.~W., Ettori S., Fabian A.~C., 2001b, MNRAS,
324, 877 
\bibitem []{} Allen S.~W., Schmidt R.W., Fabian A.~C., 2001c, MNRAS,
328, L37 

\bibitem []{} Arnaud K.~A., 1996, in Astronomical Data Analysis
  Software and Systems V, eds. Jacoby G., Barns J., ASP Conf. Ser., 101, 17

\bibitem []{} Balusinska-Church M., McCammon D., 1992, ApJ, 400, 699

\bibitem []{} Bertschinger E., Meiksin A., 1986, ApJ, 306, L1
\bibitem []{} Binney J., Cowie L. L., 1981, ApJ, 247, 464

\bibitem []{} B\"ohringer H., Fabian A.~C., 1989, MNRAS, 237, 1147
\bibitem []{} B\"ohringer H., Matsushita K., Churazov E., Ikebe Y., Chen Y., 2002, A\&A, 382, 804




\bibitem []{} Bregman J.~N., David L.~P., 1988, ApJ, 326, 639
\bibitem []{} Br\"uggen M., Kaiser C.~R., 2001, MNRAS, 325, 676

\bibitem []{} Cardiel N., Gorgas J., Aragon-Salamanca A., 1998, MNRAS,
  298, 977
\bibitem []{} Chandran B.~D.~G., Cowley S.~C., 1998, Phys. Rev. Lett.,
  80, 3077 (CC)
\bibitem []{} Churazov E., Br\"uggen M., Kaiser C.~R., B\"ohringer H.,
    Forman W., 2001, ApJ, 554, 261
\bibitem []{} Cowie L.~L., McKee  C.~F., 1977, ApJ, 211, 135

\bibitem []{} Crawford C.~S., Allen S.~W., Ebeling H., Edge A.~C.,
   Fabian A.~C., 1999, MNRAS, 306, 857
\bibitem []{} David L.~P., Nulsen P.~E.~J., McNamara B.~R., Forman W.,
   Jones C., Ponman T., Robertson B., Wise M., 2001, ApJ, 557, 546
\bibitem []{} Edge A.~C., 2001, MNRAS, 328, 762
\bibitem []{} Ettori S., Fabian A.~C., 2000, MNRAS, 317, L57

\bibitem []{} Ettori S., Fabian A.~C., Allen S.W., Johnstone R.M.,
  2002, MNRAS, 331, 635



\bibitem []{} Fabian A.~C., 1994, ARAA, 32, 227
\bibitem []{} Fabian A.~C., Daines S.~J., 1991, MNRAS, 252, 17

\bibitem []{} Fabian A.~C., Canizares C.~R., B\"ohringer H., 1994,
  ApJ, 425, 40

\bibitem []{} Fabian A.~C., Mushotzky R.~F., Nulsen P.~E~J., Peterson J.~R., 2001, MNRAS, 321, L20

\bibitem []{} Fabian A.~C., Allen S.~W., Crawford C.~S., Johnstone
  R.~M., Morris R.~G., Sanders J.~S., Schmidt R.~W., 2002, MNRAS, in press







\bibitem []{} Friaca A.~C.~S., 1986, A\&A, 164, 6

\bibitem []{} Gruzinov A., 2002, astro-ph/0203031

\bibitem []{} Heckman T.~M., Baum S.~A., van Breugel W.~J.~M.,
   McCarthy P., 1989, ApJ, 338, 48
\bibitem []{} Johnstone R.~M., Fabian A.~C., Nulsen P.~E.~J., 1987, MNRAS,
    224, 75 
\bibitem []{} Johnstone R.~M.,  Allen S.W., Fabian A.~C., Sanders J.S.,
2002, MNRAS, submitted. astro-ph/0202071
\bibitem []{} Kaastra J.~S., Mewe R., 1993, Legacy, 3, HEASARC, NASA

\bibitem []{} Kaastra J.~S., Ferrigno C., Tamura T., Paerels F.~B.~S.,
   Peterson J.~R., Mittaz J.~P.~D., 2001, A\&A, 365, L99
\bibitem []{} Malyshkin L., 2001, ApJ, 554, 561
\bibitem []{} Markevitch M. et al., 2000, ApJ, 541, 542
\bibitem []{} Mazzotta P., Kaastra J.~S., Paerels F.~B., Ferrigno C.,
  Colafrancesco S., Mewe R., Forman W.~R., 2002, ApJ, 567, L37

\bibitem []{} McNamara et al., 2000, ApJ, 534, L135

\bibitem []{} McNamara B.~R. et al., 2001, ApJ, 562, 149    
\bibitem []{} Molendi S., Pizzolato F., 2001,  ApJ, 560, 194
\bibitem []{} Motl P.~M., Burns J.~O., Loken C., Norman M.~L., Bryan G., 2001, AAS, 199, 6906



\bibitem []{} Narayan R., Medvedev M.~V., 2001, ApJ, 562, L129

    





\bibitem []{} Peres C.~B., Fabian A.~C., Edge A.~C., Allen S.~W.,
  Johnstone R.~M., White D.~A., 1998, MNRAS, 298, 416

\bibitem []{} Peterson J.~R., 2002, astro-ph/0202108
\bibitem []{} Peterson J.~R. et al., 2001, A\&A, 365, L104


\bibitem []{} Pistinner S., Levinson A., Eichler D., 1996, ApJ, 467, 162
\bibitem []{} Saito R., Shigeyama T., 1999, ApJ, 519, 48
\bibitem []{} Schmidt R.~W., Allen S.~W., Fabian A.~C., 2001, MNRAS, 327, 1057
\bibitem []{} Sparks W.~B., 1992, ApJ, 399, 66
\bibitem []{} Spitzer L., 1962, Physics of Fully Ionized Gases, New
  York: Wiley-Interscience
\bibitem []{} Spitzer L., H\"arm R., 1953, Phys. Rev., 89, 5
\bibitem []{} Stewart G.~C., Fabian A.~C., Jones C., Forman W., 1984,
  ApJ, 285, 1
\bibitem []{} Suginohara T., Ostriker J.~P., ApJ, 1998, 507, 16



\bibitem []{} Takahara M., Takahara F., 1979, Prog. Theor. Phys., 62, 1253

\bibitem []{} Tamura T. et al., 2001, A\&A, 365, L87
\bibitem []{} Tucker W.H., Rosner R., 1983, 267, 547
\bibitem []{} Vikhlinin A., Markevitch M., Murray S.~S., 2001, ApJ,
  551, 160

\end{thebibliography}
\end{document}